\documentclass[traditabstract]{aa}
\usepackage{natbib,graphicx,amsmath,amsfonts,amssymb,bm,txfonts,pstricks}

\newcommand{\dd}{\mathrm{d}}
\def\mcfost{{\sf MCFOST}}

\newcommand{\tf}[2]{\ensuremath{#1/#2}}

\def\vgtheta{v_{\mathrm{g}\theta}}
\def\cs{c_{\mathrm{s}}}
\def\gm{\mathcal{G}M}

\begin{document}

\title{Diversity in the outcome of dust radial drift in protoplanetary discs}

\author{ C. Pinte\inst{1,2} \and G. Laibe\inst{3,4}}

\offprints{C.Pinte \\ \email{christophe.pinte@obs.ujf-grenoble.fr}}

\institute{
UMI-FCA, CNRS/INSU France (UMI 3386), and Departamento de Astronom\'ia,
Universidad de Chile, Santiago, Chile
\and
UJF-Grenoble 1 / CNRS-INSU, Institut de Plan\'etologie et d'Astrophysique de Grenoble (IPAG) UMR 5274, Grenoble, F-38041, France
\and
Centre for Stellar and Planetary Astrophysics,
School of Mathematical Sciences, Monash University, Clayton Vic 3168, Australia
\and
School of Physics and Astronomy, University of Saint Andrews, North Haugh, St Andrews, Fife KY16 9SS
}
\date{Received ??; Accepted ??}

\abstract
{The growth of dust particles into planet embryos needs to
 circumvent the ``radial-drift barrier'', \emph{i.e.} the accretion of
 dust particles onto the central star by radial migration.
The outcome of the dust radial migration is governed by
simple criteria between the dust-to-gas ratio and the exponents $p$ and $q$ of
the surface density and temperature power laws.
The transfer of radiation provides an additional constraint between these
quantities because the disc thermal structure is fixed by the dust spatial
distribution.
To assess which discs are primarily affected by the
radial-drift barrier, we used the radiative
transfer code \mcfost\
to compute the temperature structure of a wide range of disc models,
stressing the particular effects of grain size distributions and
vertical settling.

We find that the outcome of the dust migration process is
  very
  sensitive to the physical conditions within the disc.
For high dust-to-gas ratios ($\gtrsim 0.01$)  and/or flattened disc structures
($H/R \lesssim 0.05$), growing dust grains can efficiently decouple
  from the gas, leading to a high concentration of grains at
a critical radius of a few AU.
Decoupling of grains from gas can occur at a large fraction ($> 0.1$) of the
initial radius of the particle, for
  a dust-to-gas ratio greater than $\approx$ 0.05.
 Dust grains that experience migration without significant growth (millimetre
 and centimetre-sized) are efficiently accreted
for discs with flat surface density profiles ($p<0.7$)  while they
always remain in the disc if the surface density is steep enough ($p>1.2$).
Between ($0.7<p<1.2$), both behaviours may occur depending on
the exact density and temperature structures of the disc.
Both the presence of large grains and vertical settling tend to favour
the accretion of non-growing dust grains onto the central object,
but it slows down the migration of growing dust grains.
If the disc has evolved into a
self-shadowed structure, the required dust-to-gas ratio for dust grains to
stop their migration at large radius become much smaller, of the order of
0.01.
All the
disc configurations are found to have favourable temperature
profiles over most of the disc to retain their
planetesimals.

}


\maketitle

\section{Introduction}
\label{sec:intro}

\begin{figure}
\includegraphics[width = \hsize]{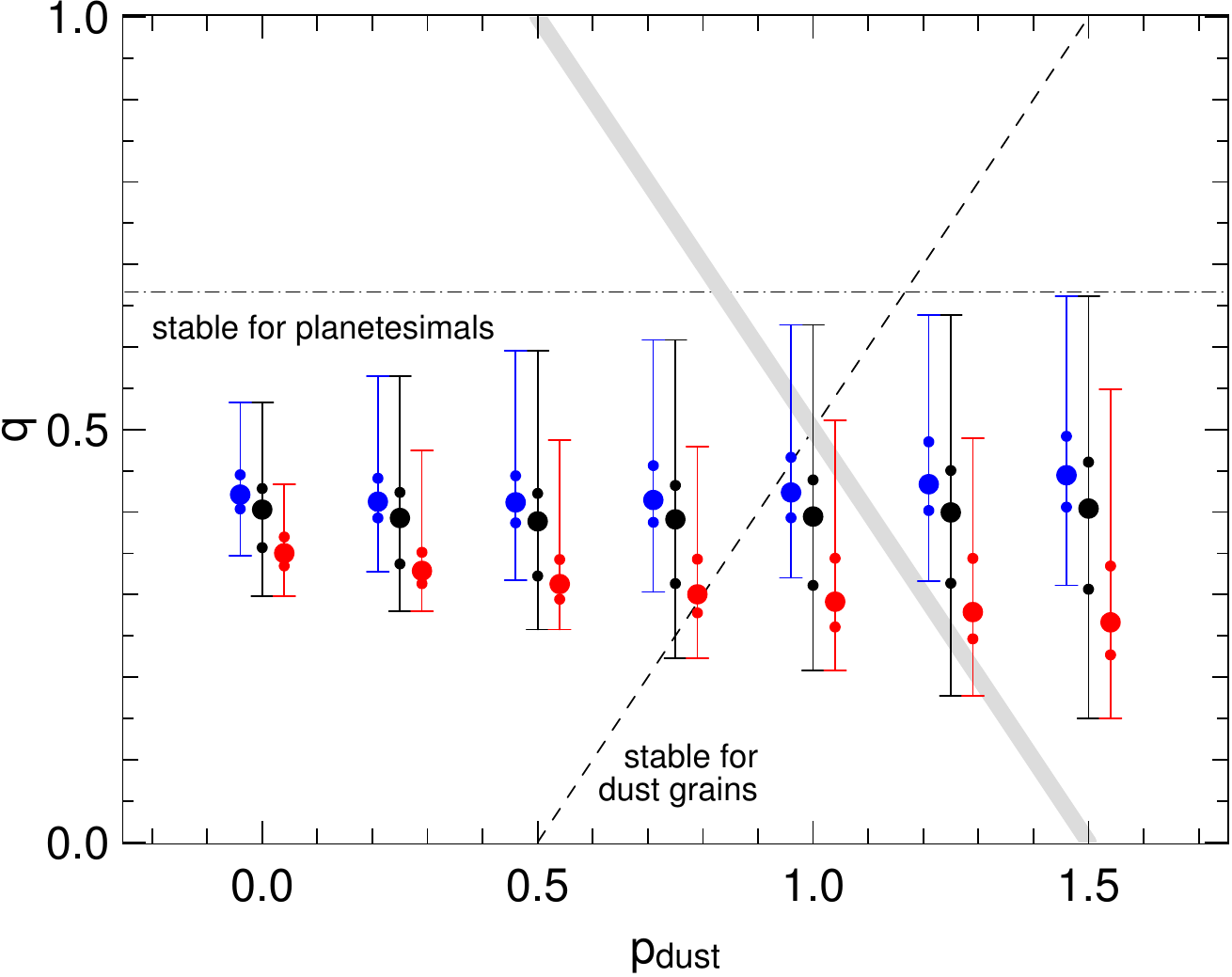}
\caption{Temperature exponent $q$ as a function of the surface density exponent
  $p_\mathrm{dust}$. Large bullets
  show the median values, smaller bullets the $1^\mathrm{st}$ and
  $2^\mathrm{nd}$ quartiles, and vertical bars the complete
  distribution of models; \emph{i.e.}, the horizontal ticks on each bar represent the
    minimum and maximum values reached in the whole set of models. Models with maximum
  grain sizes ranging from $1\,\mu$m to $10\,\mu$m are shown in red, while models
  with  $a_\mathrm{max}$ ranging from $100\,\mu$m to $10$\,mm are shown in blue. The
  different models are
  slightly shifted along the $p$ axis for clarity.  The
  plane is divided by the stability criteria for dust grains:
  $q = 1$ with growth, $-p_\mathrm{gas}+q+1/2 = 0$ (assuming
  $p_\mathrm{gas} = p_\mathrm{dust}$, black dashed line) without growth and for planetesimals: $q = 2/3$
  (black dot-dashed lines). The thick gray line indicates the
    location of steady-state, constant $\alpha$  viscous disc
    models $p_{\rm gas} = 3/2 - q$.
\label{fig:grain_size}
}
\end{figure}

\defcitealias{Laibe08}{LGFM08}
\defcitealias{Laibe2012}{LGM12}
\defcitealias{Laibe2014b}{L14}

The evolution of dust particles embedded in circumstellar discs
constitutes the first step towards planet formation
\citep{bw08,Chiang2010}. Micrometre-sized dust grains collide and grow
by coagulation, forming larger pebbles (millimetre- and
centimetre-sized, \citealt{Beckwith00,Dominik06PPV} and references
therein) that may ultimately give birth
to kilometre-sized planetesimals. Dust grains also settle towards the
disc midplane and migrate inwards as a result of the conjugate actions
of stellar gravity, gas drag, and radial pressure gradient
\citep[e.g.][herafter \citetalias{Laibe2012}]{Weidendust1977,Nakagawa86,Fromang06,Laibe2012}.
This
results in a stratified disc structure with a population of small
(micrometre-sized) grains remaining close to the surface and larger grains
located deeper
in the disc \citep[e.g.][]{Dullemond04,Duchene04,DAlessio06,Pinte07,Pinte08b}. However, the
interplay between these two mechanisms remains unclear. For
instance, how dust grains overcome the ``radial-drift
barrier'' --- \emph{i.e.} the accretion of dust grains onto the
central star on timescales shorter than the disc's lifetime. Several
mechanisms have been invoked to overcome this barrier,
such as grain growth  (\citealp[e.g.\
][hereafter \citetalias{Laibe08}]{Laibe08},
\citealp{Brauer08,Okuzumi2009,Birnstiel09,Zsom2010,Windmark2012,Garaud2013}) or trapping in pressure maxima
\citep[e.g.\ ][]{Pinilla2012,Gibbons2012}.

Another mechanism has been studied by \cite{Stepinski97,Youdin2002} and \cite{Youdin2004} and revived by \citetalias{Laibe2012}, \cite{Laibe2014a}, and \citet[][hereafter \citetalias{Laibe2014b}]{Laibe2014b} for
  growing dust grains. As the
grains move inward, their radial motion is affected by the increasing drag force,
the increasing pressure gradient, and eventually a larger size due to grain
growth. The combination of these effects can lead to a variety of outcomes. In particular,
for specific values of the exponents of
the gas surface density ($\Sigma(r) \propto r^{-p}$) and midplane
temperature ($T(r) \propto r^{-q}$),  the time required for grains to
reach smaller and smaller radii increases drastically\footnote{in the case
  where the gas accretion velocity is neglected}.
This causes dust grains to decouple from the gas phase and ``pile-up'' in the dense, inner
regions of the disc:
the migration
timescale becomes so long that dust particles are virtually stopped
and never end up accreted onto the central object.

Growing grains may also break through the drift barrier because of
   the transition from Epstein to Stokes drag at high gas densities
   \citep[][see their Fig. 11]{Birnstiel10}.

Although the kinematics of the radial
motion of the grains depends on the grain size, the outcome of the
radial drift motion essentially depends on the values of $p$ and
$q$.
\citetalias{Laibe2012} and \citetalias{Laibe2014b}
derived three analytic criteria that predict whether dust grains are ultimately accreted
onto the central star or not.  In order for dust to pile up, the
criteria are

(i) $q < 1$ and $\Lambda (r) > 1-q$ for growing grains, where
\begin{equation}
\Lambda (r) = \frac{\epsilon_{0}}{\left( p + q/2 + 3/2\right)\left(H(r)/r \right)^{2} }
\end{equation}
measures the relative efficiency between growth and migration,
and $\epsilon_{0}$ is the disc's initial dust-to-gas ratio.

(ii) $-p+q+1/2<0$ for non-growing grains in the
Epstein drag regime (which corresponds to dust grains of size smaller than
$\approx$\,10\,cm for density conditions encountered in protoplanetary discs),
and

(iii) $q<2/3$ for planetesimals ($>\,1$\,m) in the Stokes drag regime.\\
In the following, we use the terms ``growing grains'', ``non-growing
  grains'', and ``planetesimals'' to refer to these three cases.

\begin{figure*}
\includegraphics[width = 0.49\hsize]{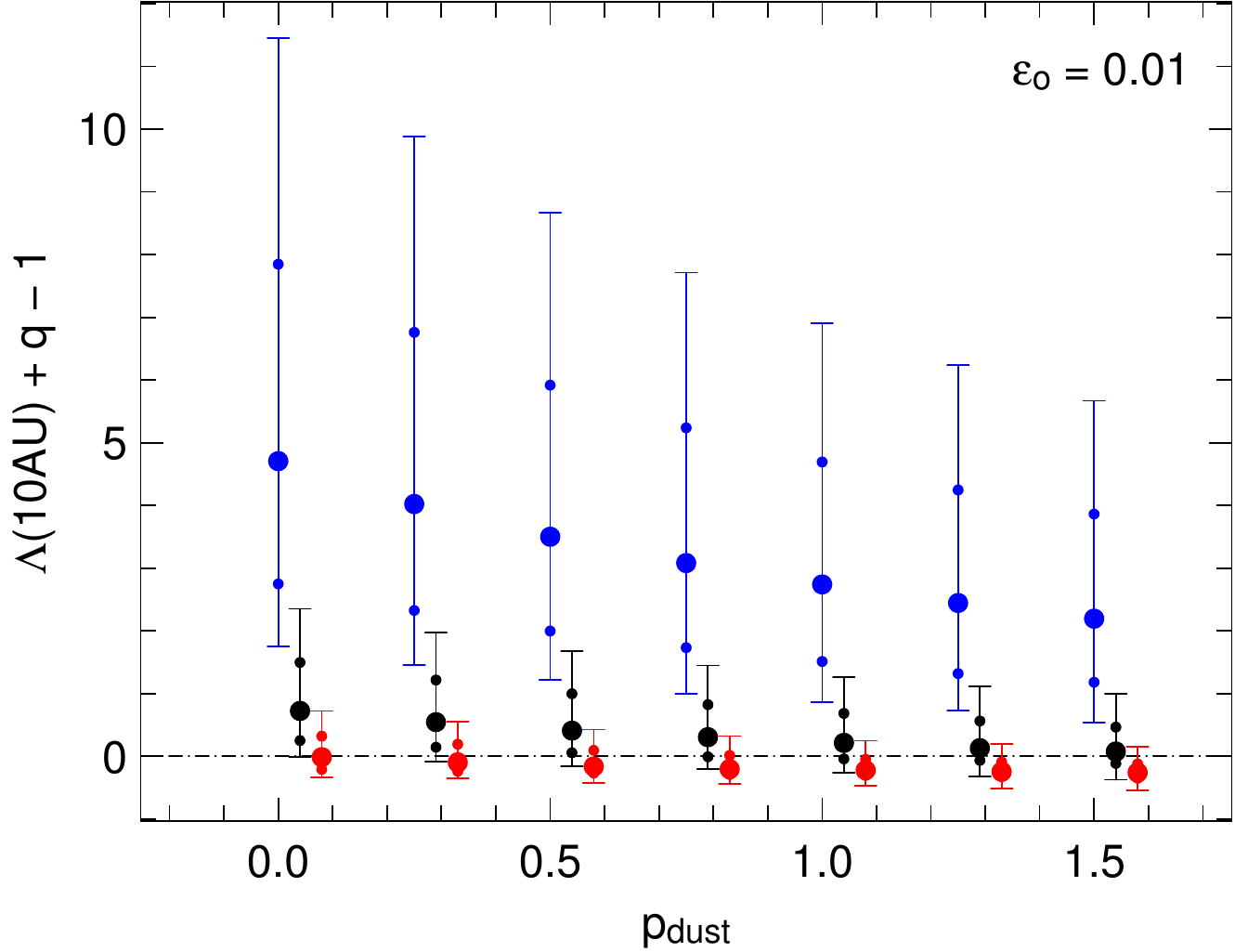}
\hfill
\includegraphics[width = 0.49\hsize]{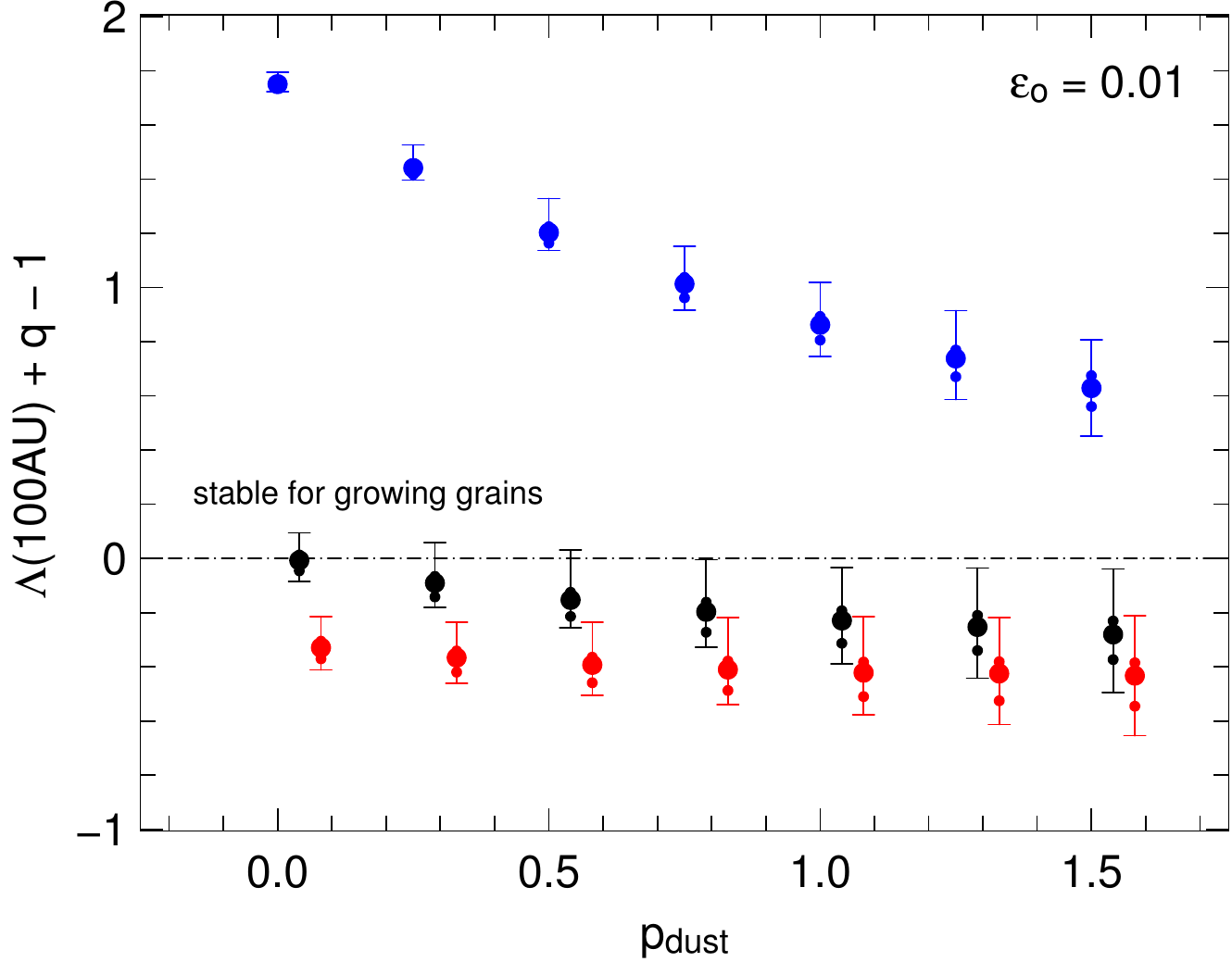}
\caption{Stability criteria $\Lambda(r) + q - 1 > 0$ for the growing grains
    as a function of the
    surface density exponent for $r = 10\,$AU (left panel) and $r = 100\,$AU
    (right panel) and an initial dust-to-gas ratio
    $\epsilon_{0} = 0.01$ in both cases. Models with scale height of 5, 10, and
    15\,AU at a radius of 100\,AU are
    shown in blue, black, and red respectively. The vertical axes are
    different in both panels. The second part of the stability criteria for
    growing grains : $q < 1$ is
    always satisfied in the disc models we have explored, see
    Fig~\ref{fig:grain_size}.
\label{fig:Lambda}}
\end{figure*}

\begin{figure*}
\includegraphics[width = 0.49\hsize]{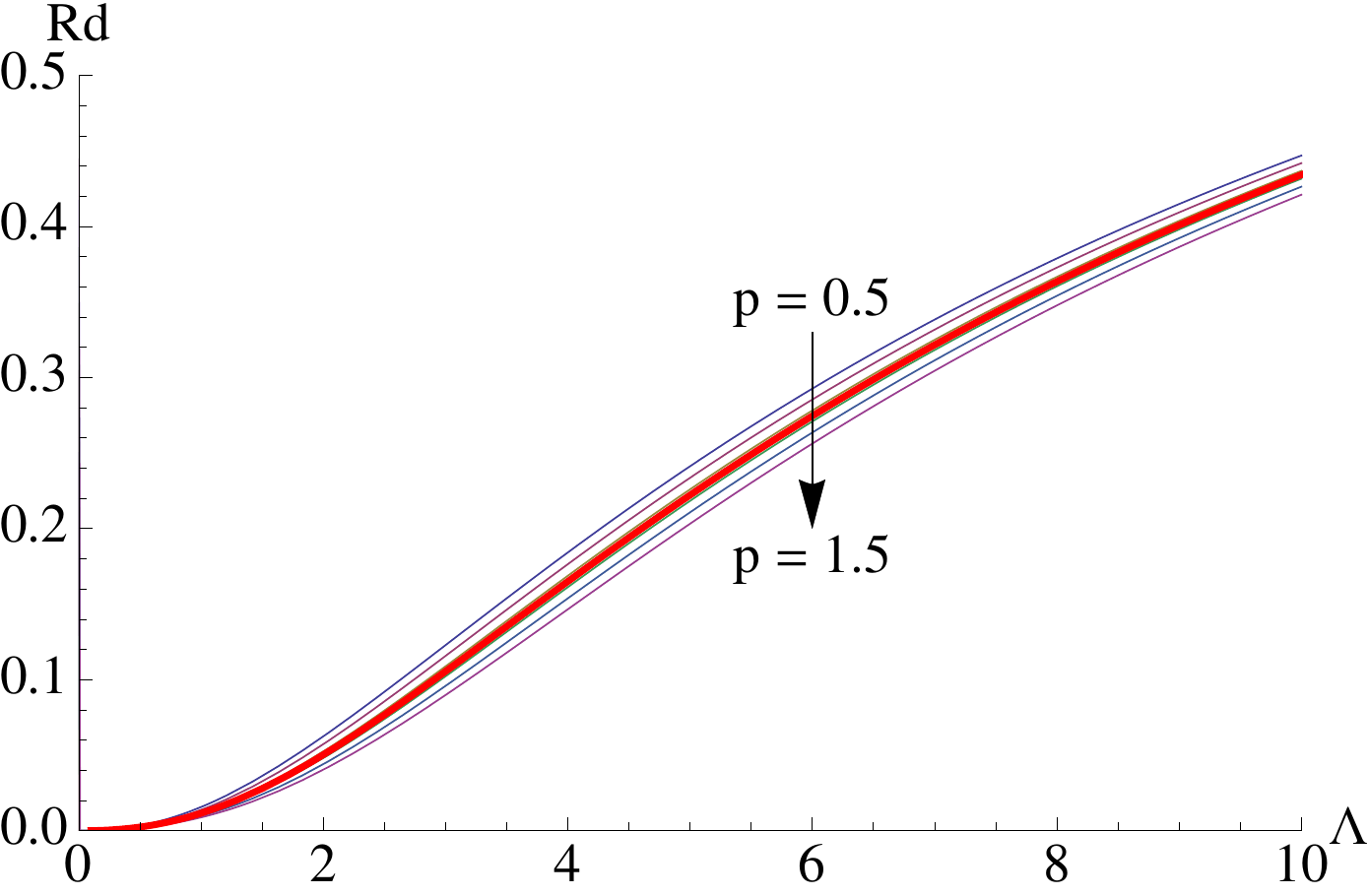}
\hfill
\includegraphics[width = 0.49\hsize]{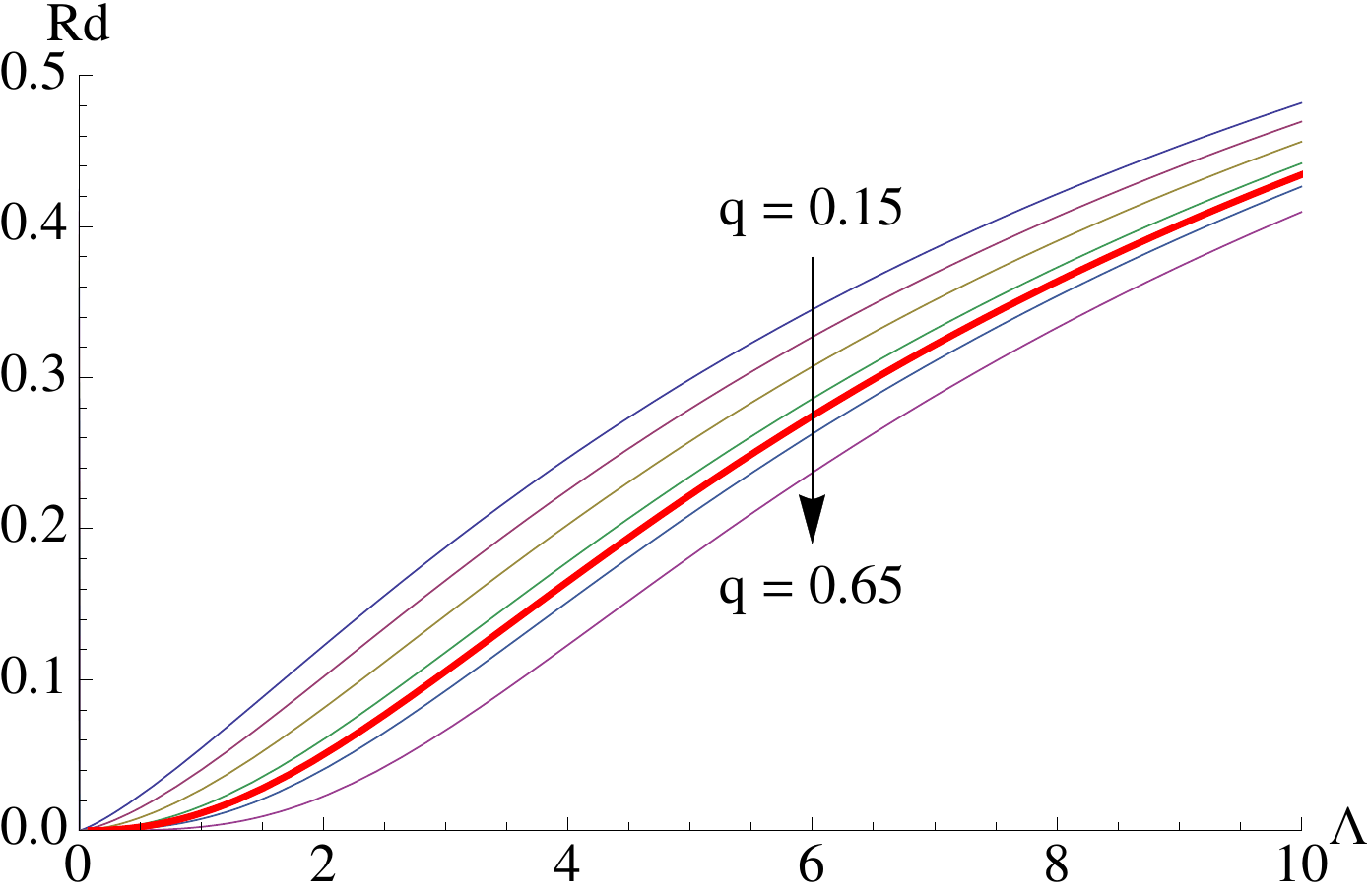}
\caption{Relative decoupling radius for growing dust grains (\emph{i.e.}
    decoupling radius divided by initial radius of the grains in the disc) as a function of $\Lambda$
for various values of $p$ (left) and $q$ (right). The red thick solid line
shows a typical Classical T Tauri Star disc with $p = 1$, $q = 0.5$.
\label{fig:Rdec}}
\end{figure*}

In these former studies, $p$ and $q$ were considered to be independent
parameters that could take any value.
However, the radial temperature profile of the disc is set
by the transfer of radiation, which depends on both the dust
properties and the disc geometry. The $q$ parameter therefore
depends on the dust surface density profile $p_\mathrm{dust}$
and can only take a limited range of
values. Furthermore, the grain size distribution
(which affects the wavelength dependence of the opacity) and vertical settling
(which lowers the $\tau=1$ surface, changing the amount of stellar
light intercepted by the disc as a function of the radius) are
expected to affect the value of $q$ and, in turn, the outcome of the
radial migration.

In this paper, we present the first results of the effect of radiative
transfer on the radial-drift barrier in both the Epstein and
Stokes regimes. Specifically, we determine i) which parts of the ($p$,
$q$) diagram are populated with physical models, ii) whether these discs
are strongly affected by the radial-drift barrier process, and iii) the
key parameters that affect the stability of the dust particles with
respect to radial migration.

\begin{table}
\begin{center}
\begin{tabular}{lcc}
\hline Parameter & Range of values explored & Step \\ \hline
$p_\mathrm{dust}$ & 0 -- 1.5 & 0.25 \\
$H_0$ (gas) [AU] at 50\,AU &  5 -- 15 & 5\\
$\beta$ (gas) &  1.0 -- 1.35 & 0.05 \\
$a_{\rm max}$ [$\mu$m]&  1 -- 10\,000 & factor 10\\
$\alpha_\mathrm{SS}$ & $10^{-4}$ -- $10^{-1}$ + (no settling)
& factor 10\\
\hline
\end{tabular}
\end{center}
\caption{Range of values explored for the various parameters in the
  grid of models computed with \mcfost.
\label{Tab:param}
}
\end{table}

\section{Methodology}
\label{sec:methodo}

In the following, we discuss the outcome of grains that are large enough
(\emph{i.e.} $> 10\,\mu$m) to be
significantly affected by radial migration. These
grains first settle toward the disc midplane before they migrate
inward. \citetalias{Laibe2012}'s
  criteria were originally derived for the migration of dust grains in
  the disc midplane  but still remain valid if the grains are located close
  to the midplane, \emph{i.e.} at lower altitudes than the local
  scale height (see App.~\ref{app:crit3D}).

The criteria defined in \citetalias{Laibe2012} and \citetalias{Laibe2014b} apply for $p$ and $q$ corresponding
to the gas phase. Here we only consider young gas-rich and radially optically thick
discs (corresponding to the classical T Tauri stage). We assume that the density  in
the disc midplane is high enough to ensure thermal
equilibrium between gas and dust and we take both T$_\mathrm{gas}$ =
T$_\mathrm{dust}$ and $q_\mathrm{gas} = q_\mathrm{dust}$, and
only use the notation $q$ in the following.
Disc temperature structures are computed for given density structures,
\emph{i.e.} for given values of $p_\mathrm{dust}$.
We assume that in the initial state of the discs, gas and dust grains
are radially  well mixed, implying $p_\mathrm{gas} =
p_\mathrm{dust}$. We discuss deviations from this case in section \ref{sec:discuss_p_dust}.

\subsection{Model description}

We assume an axisymmetric flared density structure with a Gaussian
vertical profile $\rho(r,z) = \rho_0(r)\,\exp(-z^2/2\,h(r)^2)$ where
$r$ is the distance from the star in the disc midplane and $z$ the
altitude above the midplane.
We use power-law distributions for the dust surface density
$\Sigma(r) = \Sigma_0\,(r/r_0)^{-p_\mathrm{dust}}$  and the scale
height $ h(r)=h_0\,(r/r_0)^{\beta}$ where $h_0$ is the scale height at
radius $r_0=100$\,AU and $\beta$ the disc flaring
  exponent\footnote{$\Sigma(r)$ and $T(r)$ as observed in discs strongly
  differ from the so-called minimal mass solar nebula disc models
  (e.g. \citealt{Weidendust1977,Hayashi1981}) used in most of theoretical
  studies.}. We fix the density structure rather than assume
  hydrostatic equilibrium in order to assess the effect of each model
  parameter individually. In particular, the flaring parameter and scale
  height are free parameters that are
 not self-consistently derived from the temperature calculation. The disc extends from an inner cylindrical radius
$R_\mathrm{in} = 0.1\,$AU to an outer limit $R_\mathrm{out} = 300\,$AU.
Since we are only interested in the disc's temperature gradient and not
its absolute value, we use a fixed central object
(T$_\mathrm{eff}=4\,000\,$K, R$= 2\,$R$_\mathrm{sun}$) and a fixed
dust disc mass of M$_\mathrm{dust} = 10^{-4}$M$_\mathrm{sun}$. The disc
mass directly influences the local
stopping time (or equivalently, the Stokes number) and thus the grains' kinematics.
However, the
Stokes number is only a scaling factor in the equations of motion of the grains,
and its value does not affect
whether the grains will pile up or decouple at a finite
radius. The key parameters are the dependencies of the Stokes number
and the Stokes number's rate with respect to both the grain size and
distance to the central star, which are independent of the disc mass/density
\citepalias{Laibe2012,Laibe2014b}. In particular, the decoupling radius
for growing dust grains is independent of the disc mass.

Dust grains are modelled as homogeneous and spherical particles with
sizes distributed according to the power law $\dd n(a) \propto
a^{-3.5}\,\dd a$ between $a_{\mathrm{min}}$ (fixed to $0.01\,\mu$m) and
$a_{\mathrm{max}}$.  Absorption and
scattering opacities, scattering phase functions, and Mueller matrices
are calculated using the Mie theory.

The thickness of the dust
layer results from the competition between gravity and turbulent
stirring and  depends on both the grain size and
the turbulent viscosity coefficient
$\alpha_{\mathrm{SS}}$ \citep{ShakuraSunyaev1973}.  Dust settling is
implemented using a diffusion-advection equation with a
variable diffusion coefficient, which has been found this to be a reasonable
approximation to global MHD simulations of stratified and turbulent discs
\citep[][Eq.~26]{Fromang09}.
 The degree of dust settling is
changed by varying $\alpha_{\mathrm{SS}}$. For comparison, models without any settling
are also included.

\subsection{Computing the stability criteria}

To ascertain the importance of radiative transfer on the outcome
of the radial migration of dust grains, we computed a grid of disc
models with the \mcfost\ radiative transfer code \citep{Pinte06,Pinte09} varying both
the dust optical properties and the disc density distribution and geometry (Table~\ref{Tab:param}). This includes the $p_\mathrm{dust}$
exponent, which is treated here as an input parameter and is not evolved
  with time.

\mcfost\ solves the continuum transfer with a Monte Carlo method. In short,
the code stochastically propagates photon packets in 3D through the
disc: the transport of packets is governed by successive scattering,
absorption, and re-emission events, which are determined by the local
dust properties. The temperature structure is calculated using the
immediate re-emission algorithm of \citet{Bjorkman01} combined with a
continuous deposit of energy to estimate the mean intensity
\citep{Lucy99}.
The resulting midplane temperature profiles in the outer parts of the
discs ($r\,\gtrsim\,10$\,AU) are fitted by a power law $T(r) \propto
r^{-q}$. Radiative transfer calculations
  have shown that this is a good approximation of the temperature profile in the
outer disc \citep[see for instance the radial temperature profiles presented
  in][]{DAlessio98,Dullemond02,Pascucci04,Pinte09}. We additionally compute the local slope of the midplane
temperature as $q_\mathrm{loc} = - \tf{\dd \log (T(r))}{\dd \log (r)}$ in
order to determine in which regions of the discs the temperature profile can be
approximated by a power law and the simple analytical
criteria can be applied, and to assess how the migration will be affected when
the temperature profiles deviate strongly from the power law.

\subsection{Model limitations}
\label{sec:limitation}
This work is based on analytical calculations of the dust
  evolution in discs, which necessarily implies some simplifying
  assumptions:
\begin{itemize}
\item The equations are mathematically valid only when the temperature profile is a
  power law. We found that this is the case for $r > 10\,$AU in most of the disc
  configurations. When the temperature profile differs from a power law, the
  general trend remains, but the predictions become more qualitative.
\item The dust-to-gas ratio is assumed to be constant when integrating the
  analytical expressions. But because dust grains migrate, this is not the case
  in discs.
\item The underlying growth model remains simple and does not consider
  more complex
  effects that have been shown to affect the evolution of the dust population
  in young discs \citep[see][for a recent review]{Dullemond13}: for instance,
  bouncing or fragmentation barrier, charging barrier or
 porosity evolution \citep{Okuzumi12}.
\item The radial gas flow, which may be important in the
very central parts of the disc was not considered here. It may also act to
remove the dust particles from the disc, preventing them from being accreted onto
planetesimals, as it may transport particles that are coupled to the
gas inwards, even when they are not drifting.
\end{itemize}

Additionally, we do not consider the case of large dust grains that are
observed at large distances from the star
  \citep[see for instance][who measured critical radii\footnote{The critical
      radius is defined as $\Sigma(r)
      \propto (r/R_c)^{-\gamma} \exp(-(r/R_c)^{2-\gamma})$, approximately
      2/3 of the mass lies within the critical radius} ranging from $\approx$ 15 to $\approx$
    200\,AU]{Hughes08,Andrews09,Andrews10b,Isella09}. The presence of
  millimetre sized grains at distance much greater than a few 10\,AU requires
  additional mechanisms, such as local pressure maxima to trap the dust grains \citep[e.g.][]{Pinilla12}.

\section{Results and discussion}
\label{sec:results}

\begin{figure*}
\includegraphics[width = 0.49\hsize]{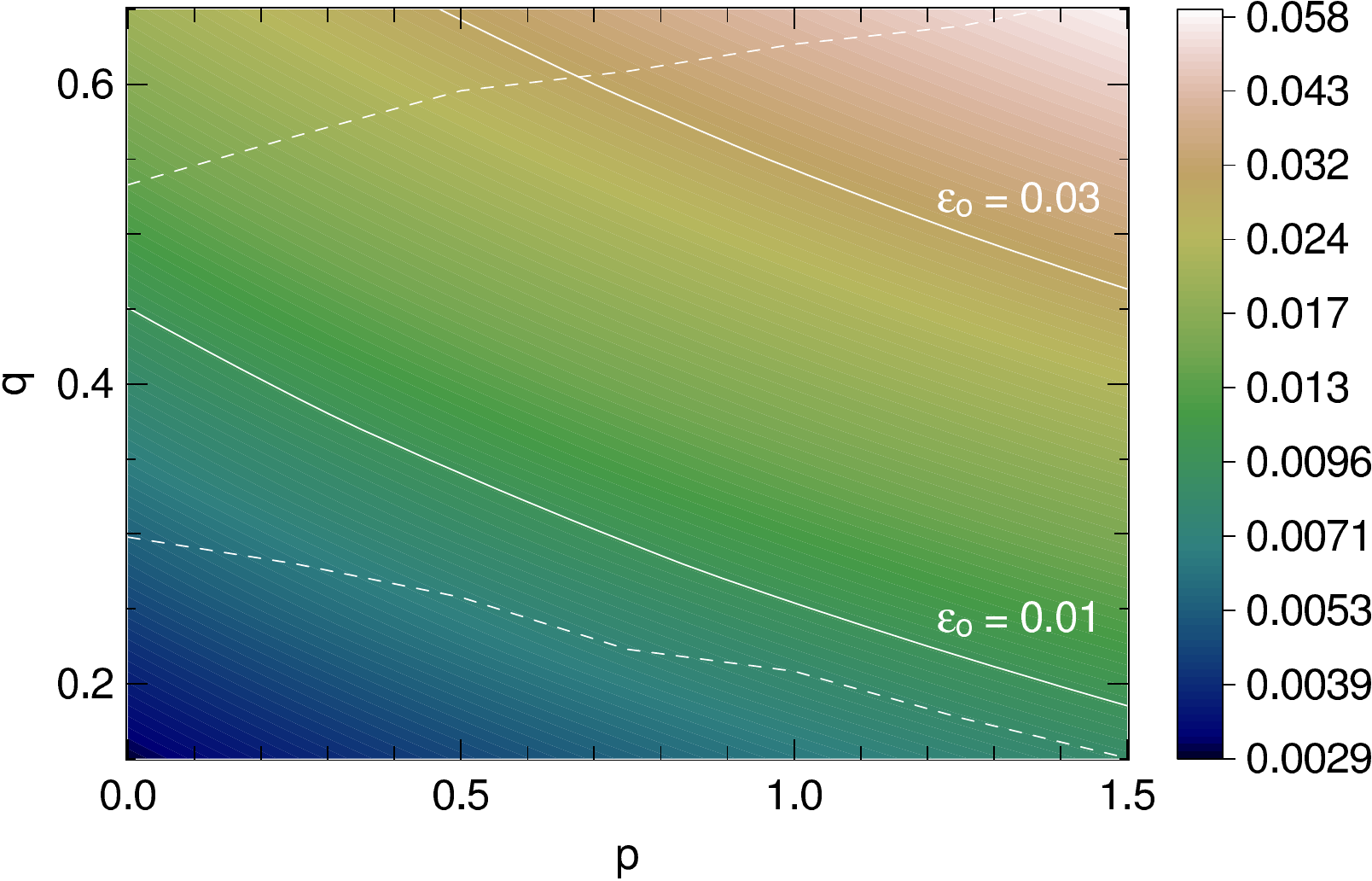}
\hfill
\includegraphics[width = 0.49\hsize]{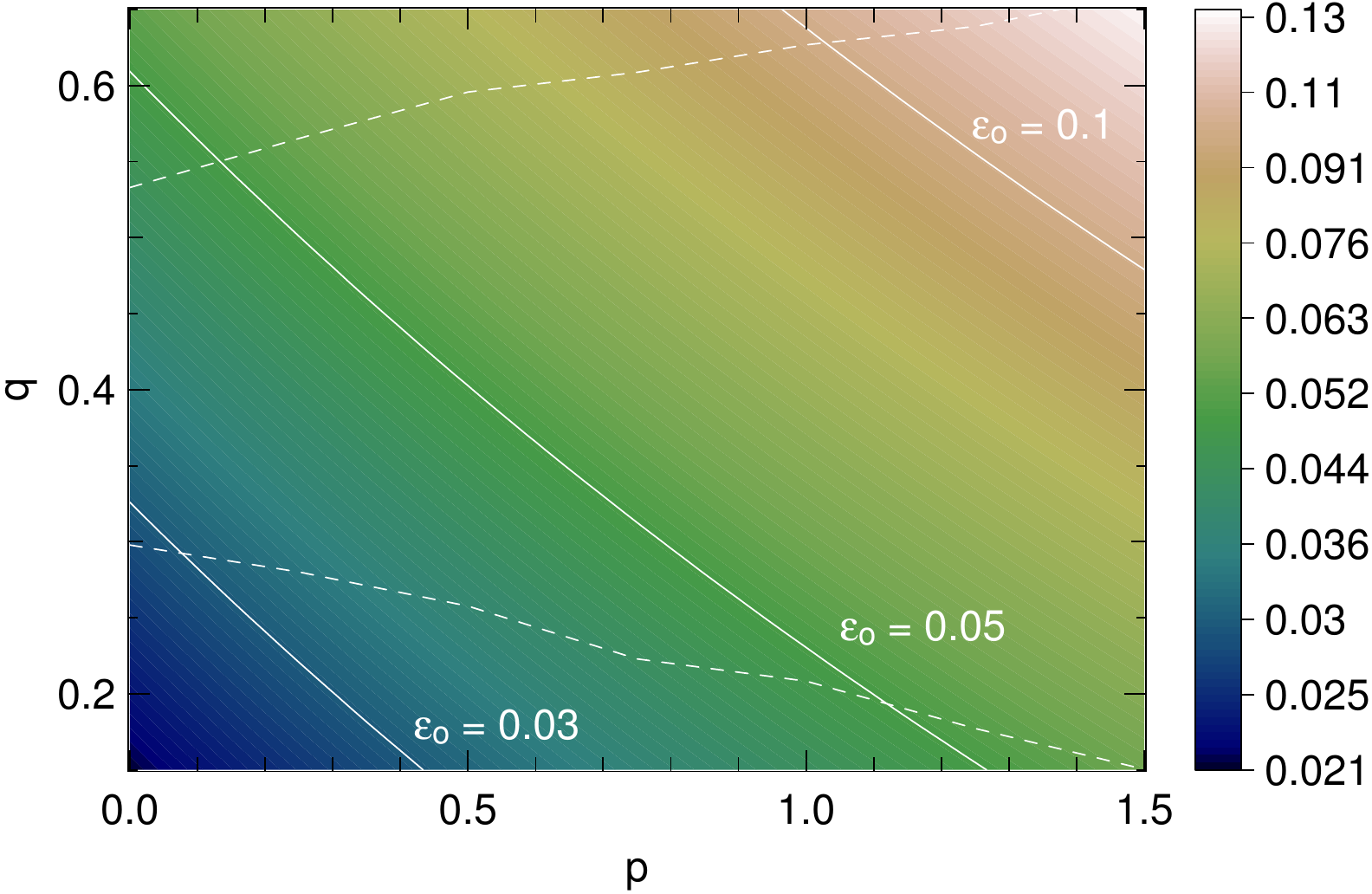}
\caption{
Initial dust-to-gas ratio required for growing dust grains to
  reach a relative decoupling radius $R_\mathrm{dec} = 0.01$ (left)
  and $0.1$ (right)
 for a scale height of 0.1 times the local radius. The ranges for $p$ and $q$
 values corresponding to the values we have in the radiative transfer
 calculation are delimited by the dashed lines
 (see Fig.~\ref{fig:grain_size}). Dust-to-gas ratio between $\approx$ 0.01 and 0.03, and
 between $\approx$ 0.05 and 0.1 are sufficient for a
 grain starting at 100\,AU to stop its migration at 1 and 10\,AU, respectively.
\label{fig:dust-to-gas}
}
\end{figure*}

The values of $q$ obtained from
radiative transfer modelling for the various disc models are presented
in Fig.~\ref{fig:grain_size}.

\subsection{Stability of growing dust grains relative to radial migration}

Initially small, interstellar-like dust grains ($\lesssim 1\,\mu$m) experience simultaneous migration and growth in the
  outer disc, until they reach a size of at least a few millimetres, when
  the sticking probabilities during collisions are strongly reduced (``bouncing
  barrier'', see section~\ref{sec:non-growing}).
For these growing dust grains, the criterion $q<1$ is satisfied for
  every disc's configuration (see Fig.~\ref{fig:grain_size}). If this first requirement were not fulfilled,
  growing grains would not pile up in discs. Figure~\ref{fig:Lambda} shows
  the values of $\Lambda(r) + q -1$ for an initial gas-dust-ratio of
  $\epsilon_{0} = 0.01$. The pile-up of growing grains mainly depends on
  the value of $\epsilon_{0}$: higher values of $\epsilon_{0}$ favour the
  pile-up, which becomes systematic for $\epsilon_{0} \gtrsim 0.02$;
  \emph{i.e.}, $\Lambda(r)$ is always larger than $q -1$.
The case  $\epsilon_{0} = 0.01$
  results in values of $\Lambda + q - 1$ close to 0, \emph{i.e.}
  at the limit of stability. In that case, the stability of growing dust grains
depends mainly on the local scale height \citep[as previously discussed
    by][see their Eq. 43]{Brauer08}: smaller scale heights tend to
  favour dust pile-up. For instance,
scale heights smaller than 10\,AU at a
reference radius of 100\,AU result in a pile-up for grains starting at
100\,AU.
$\Lambda(r)$ increases as $r$ decreases and the conditions for pile-up
become less stringent as smaller radii, see for instance the right-hand panel of
Fig.~\ref{fig:Lambda} at $r =10$\,AU, where most of the disc configurations
will be favourable for dust grain pile-up.

The previously described criteria only tell us whether pile-up will occur. For
  this pile-up to help the formation of planetary cores, it has to occur at a
  distance that is sufficiently far from the star that the dust grains do not
  reach the very central regions where dust sublimation or the gas accretion
  flow will remove the dust particles before they can be incorporated in planetesimals.
The pile-up of growing dust grains, when it happens, is extremely efficient (the migration
  time goes as an exponential of a power law with respect to the grain's
  position), and the grains literally stop their migration. The stopping radius
  of the grains depends on the relative efficiency of the growth and migrations
  processes and the value of $\Lambda$. Figure~\ref{fig:Rdec} shows the
  decoupling radius $R_{\mathrm{d}}$ as a fraction of the initial radius and as a function of the
  $\Lambda$ parameter for the various values of $p$ and $q$ obtained in our
  grid of radiative transfer models. We estimated $R_{\mathrm{d}}$ numerically,
  by solving simultaneously  the equations for grains migration and growth using the (YL; Vnew) model of \citetalias{Laibe2014b}. We calculated $R_{\mathrm{d}}$ as
the grain's radius in the asymptotic limit. We used $t =
  10^{4}$\,years here and checked that the asymptotic limit was reached.
The decoupling radius is essentially not sensitive to the value of the surface
density exponent $p$. However, a noticeable dispersion is obtained when varying
the temperature exponent $q$. The decoupling of the grains is more efficient
for discs with smooth thermal profiles.
For $\Lambda = 1$, the decoupling radius
  can be located at up to 6\,\%  of
  the initial distance to the central star for low values of $q$ (0.15). For
  larger $q$, the corresponding decoupling radius decreases and reach values
  lower than 1\,\% for $q=0.65$. For $\Lambda = 5$, the decoupling occurs at
  values ranging between 18 and 30\,\% of the initial radius.
This process was also observed in numerical simulations as
presented in \citetalias{Laibe08}, explaining why dust grains can be found to overcome
the radial-drift barrier, depending on the physical conditions in the disc.

Figure~\ref{fig:dust-to-gas} presents the dust-to-gas ratios that are required
for the dust grains to decouple at a final radius equal to 1\,\% (left) and
10\,\% (right) of their initial radius. Dust grains starting a few hundred AU
will stop their migration at a distance greater than a few AU as long as the dust-to-gas ratio is higher than
the generally assumed interstellar value $0.01$. A slight enhancement compared
to this value, for instance $\epsilon_\circ = 0.03$, is enough to obtain a
decoupling in almost all the disc configurations we have explored.
For a higher dust-to-gas ratio, the decoupling will occur at even greater
distances. With $\epsilon_\circ$ between $0.05$ and $0.1$, most of the disc
configurations will show dust grains decoupling at a radius larger than 10\,\%
of their initial radius.
The required $\epsilon_\circ$ for dust grains to pile up at a given radius is
smaller for discs with flatter surface density profiles and shallower midplane
temperature gradients. For a given surface density, there are large variations
in the required $\epsilon_\circ$ depending on the temperature structure of the
disc. For instance, for $p = 0.75$, the required values for $\epsilon_\circ$ to
obtain dust pile-up
range from $7.5\times10^{-3}$ to 0.032 for a relative decoupling radius of 0.01 and from
0.043 to 0.085  for a relative decoupling radius of 0.1.
 Taking
the details of the temperature structure and radiative transfer effects into
account is therefore
critical in dust evolution simulations to accurately predict the outcome of the
dust migration processes.

\subsection{Stability of non-growing dust grains and planetesimals relative to
  radial migration}
\label{sec:non-growing}

When dust grains reach centimetric size, they start bouncing
  instead of sticking during collisions. This reduces the growth efficiency
  dramatically \citep{Guttler2010,Zsom2010,Windmark2012}, though stochastic
  processes may assist the formation of larger bodies
 (\citealp{Okuzumi11,Windmark2012b}, but see also their Corrigendum
    \citealp{Windmark2012c}, and \citealp{Garaud2013}).
 When the growth becomes negligible, the migration outcome is given by the criterion derived in \citetalias{Laibe2012}:
 $-p_\mathrm{gas}+q+1/2<0$ (displayed in Fig.~\ref{fig:grain_size}). We note that all discs
with shallow initial profiles ($p_\mathrm{dust}\lesssim0.7$) will lose
a large fraction of their grains by accretion onto the central
star.
In the absence of additional mechanisms to slow down the migration, and
if these grains do not cross regions where the conditions allow them to grow
again, these discs will
be efficiently depleted in millimetre-size grains.
 For profiles with $p_\mathrm{dust}>1.2$, grains pile up and the discs
 efficiently retain their grain population in the inner disc. For intermediate surface
 density profiles ($0.7<p_\mathrm{dust}<1.2$),
 either dust migration or dust pile-up can happen, depending
 on the details of the disc structure and grain properties.
Interestingly, this includes the
 standard case $p_\mathrm{gas} = 1$, which is the limiting profile of a viscous
 evolution after several viscous timescales (for a
constant $\alpha_\mathrm{SS}$). This range of values for
 $p_\mathrm{dust}$ is close
 to the observed surface density in the millimetre regime.
For instance, \citet{Andrews2010} found a narrow distribution of surface
density gradients for a sample of discs in Ophiuchus: $\left<p_\mathrm{dust}\right> \simeq
0.9 \pm 0.2$, which does not seem to depend on the millimetre flux (in
their Table\,4, only one disc in 12 is found to have
$p_\mathrm{dust}<0.7$), suggesting that most of the discs are located at the
interface between the stable and unstable
regimes with respect to the migration of their millimetre/centimetre grains.

The highest value of $q$ we obtain remains less than the critical
value of $2/3$ predicted for planetesimals to be efficiently
accreted onto the central star. The thermal profile of the outer disc is
therefore favourable to retaining potential planetesimals, regardless
of the grain distribution.

For non-growing grains or planetesimals, the pile-up
  process is
  smoother than for the growing grain case. There is no actual stopping point since
  the pile-up occurs progressively, slowing down the particles more and more as
  they reach the disc's inner region. (The migration time goes as a power law
  with respect to the grain's position.) Quantitatively, particles reduce their
  migration efficiency compared to the case without pile-up when they reach a
  radius that is roughly 10$\%$ of their initial distance to the central star
  (Fig.~3 of \citetalias{Laibe2012}). Then, it takes an order of magnitude more
  time for the grain
  to decrease its radial position by another factor 10.

For instance, in a disc with $p=0.4$ and $q=0.4$ (no pile-up) and a disc
  mass of 0.1\,M$_\odot$ and gas-to-dust ratio of 100, it takes $\simeq
3.8 \times10^{5}$\,years
  for a $1$\,mm grain initially located at $100$\,AU to reach $R=10$\,AU, a
  mere additional $\simeq 3 \times10^{4}$\,years to reach $R=1$\,AU and an even
  fewer additional $\simeq 4 \times10^{3}$\,years to reach $R=0.1$\,AU  (\emph{i.e.} $\simeq 4.1\times10^{5}$\,years in total).
 For comparison, in the same disc but with $p=1.6$ and $q=0.4$ (pile-up), it takes $\simeq 3.9\times10^{5}$\,years
  for a $1$\,mm grain initially located at $100$\,AU to reach $R=10$\,AU (\emph{i.e.} roughly the same time as without the pile-up) but more additional $\simeq 1.8 \times10^{6}$\,years to reach $R=1$\,AU and an even more additional $\simeq 9.1 \times10^{6}$\,years to reach $R=0.1$\,AU (\emph{i.e.} $\simeq 1.1\times10^{7}$\,years in total). Figures~6 to 10 of \citetalias{Laibe2012} shows a parameter study for the time required for a grain to reach the inner regions of
typical classical T-Tauri discs. The global behaviour is not modified if
  the disc mass is changed and only the respective times are scaled. For a lower
disc mass, the corresponding times are reduced in the case of pile-up, while
they are increased when there is no pile-up.

\begin{figure}
\includegraphics[width = \hsize]{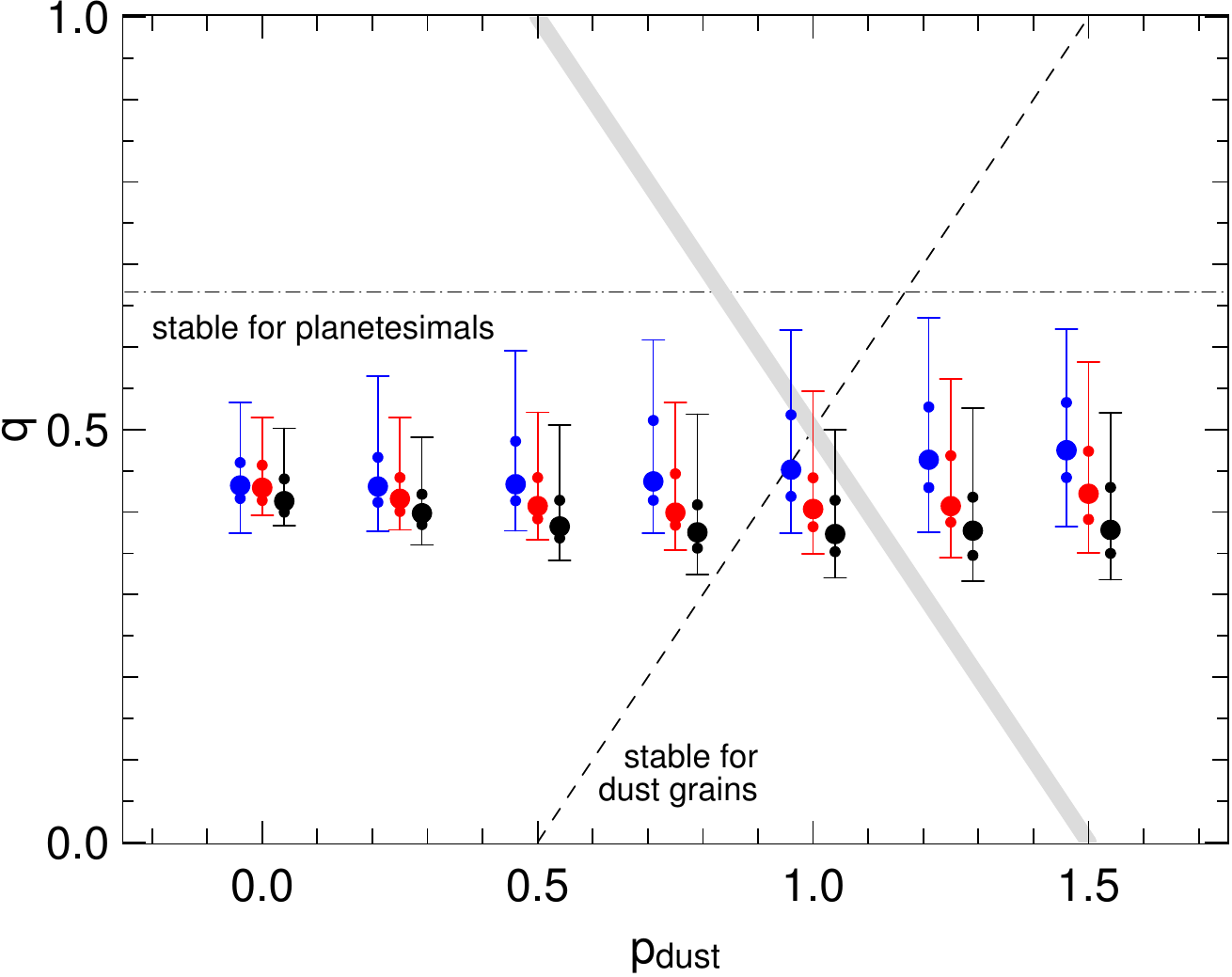}
\caption{Same as Fig.~\ref{fig:grain_size} but considering only models with
  $a_{\textrm{max}}= 10\,$mm. Models are separated according to their
  degree of settling: no
  settling (black), low
  settling ($\alpha_\mathrm{SS} \geq 10^{-2}$, red), or
  strong settling ($\alpha_\mathrm{SS} < 10^{-2}$,
  blue).
\label{fig:settling}
}
\end{figure}

\subsection{Effect of grain size distribution \& vertical settling}
\label{sec:grain_size}

In the case of optically thin dust with  an opacity law that can be
described by a  power law $\kappa_\textrm{abs} \propto
\lambda^{-\beta_\mathrm{dust}}$, the temperature is determined by
\begin{equation}
4\pi\,\int_0^\infty
\kappa_\circ\left(\frac{\lambda}{\lambda_\circ}\right)^{-\beta_\mathrm{dust}}\,
B_\lambda(T(r))\,\dd\lambda = E_\mathrm{abs}(r) \propto \frac{E_*}{r^2},
\end{equation}
resulting in a temperature profile  $T(r) \propto r^{-\frac{2}{4+
\beta_\mathrm{dust}}}$. As the effective grain size grows,
$\beta_\mathrm{dust}$ varies from 2 for small ISM-like grains to 0 for
millimetre or larger grains, the
resulting $q$ values range from $1/3$ to $1/2$.
Taking the details of radiative transfer into account, \emph{e.g.}
high optical depth and shadowing effects, we obtain values of $q$
ranging from 0.10 to almost 0.65 with the same trend of higher $q$
values for larger grains, on average.
(Fig.~\ref{fig:grain_size} illustrates this effect with 2
subsets of the disc models:
$a_{\mathrm{max}} \leq 10\,\mu$m and
and $a_{\mathrm{max}} \geq 100\,\mu$m.)
As a consequence, discs where dust grain have grown will
have a steeper temperature profile and will be more prone to losing their grain
population by accretion onto the central object.
In contrast, a possible regeneration of micrometre-sized grains
\citep{Brauer2008,Guttler2010,Zsom2010} will make the temperature
profile shallow and stabilise the outcome of the millimetre grains.

\begin{figure*}
\includegraphics[width = 0.49\hsize]{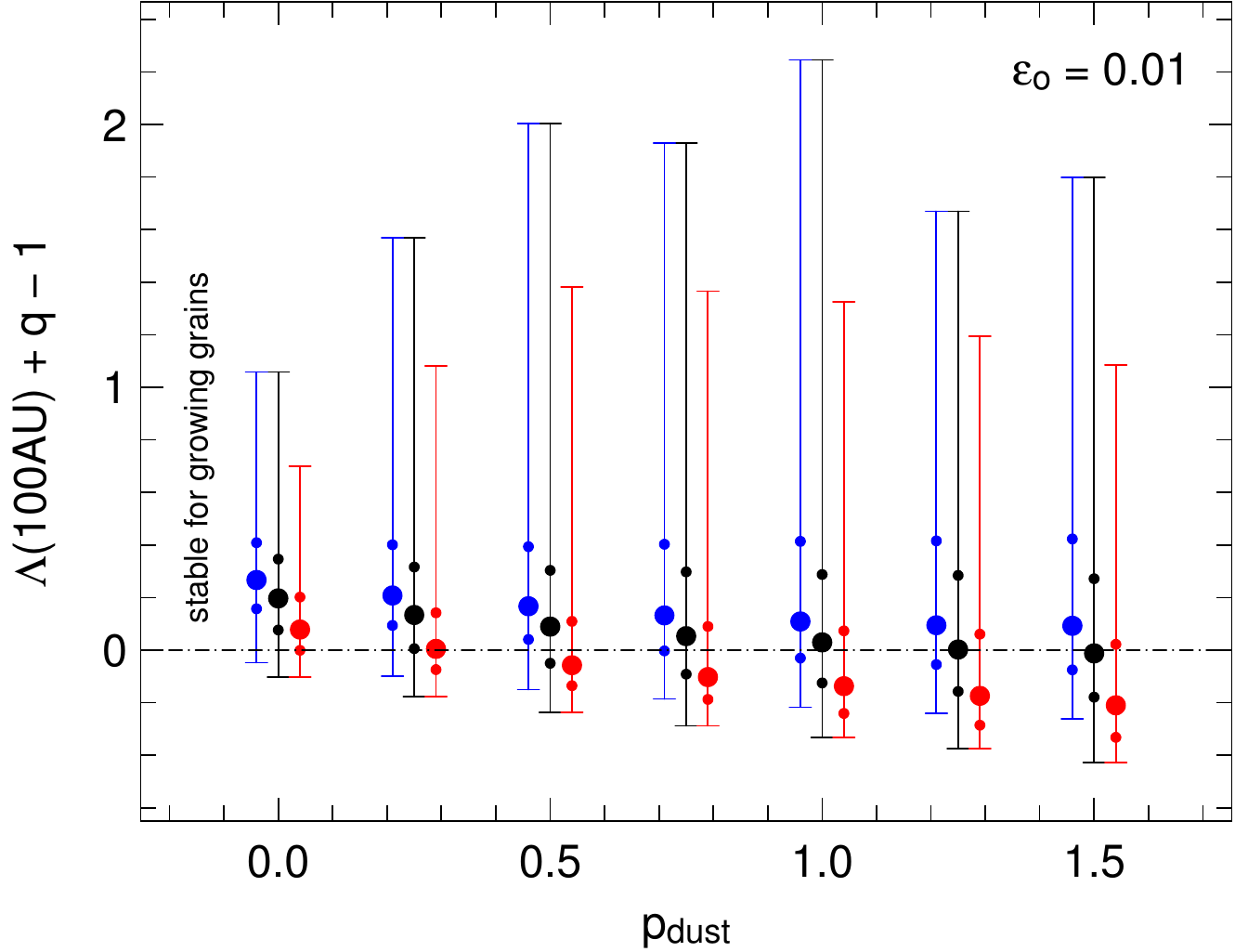}
\hfill
\includegraphics[width = 0.49\hsize]{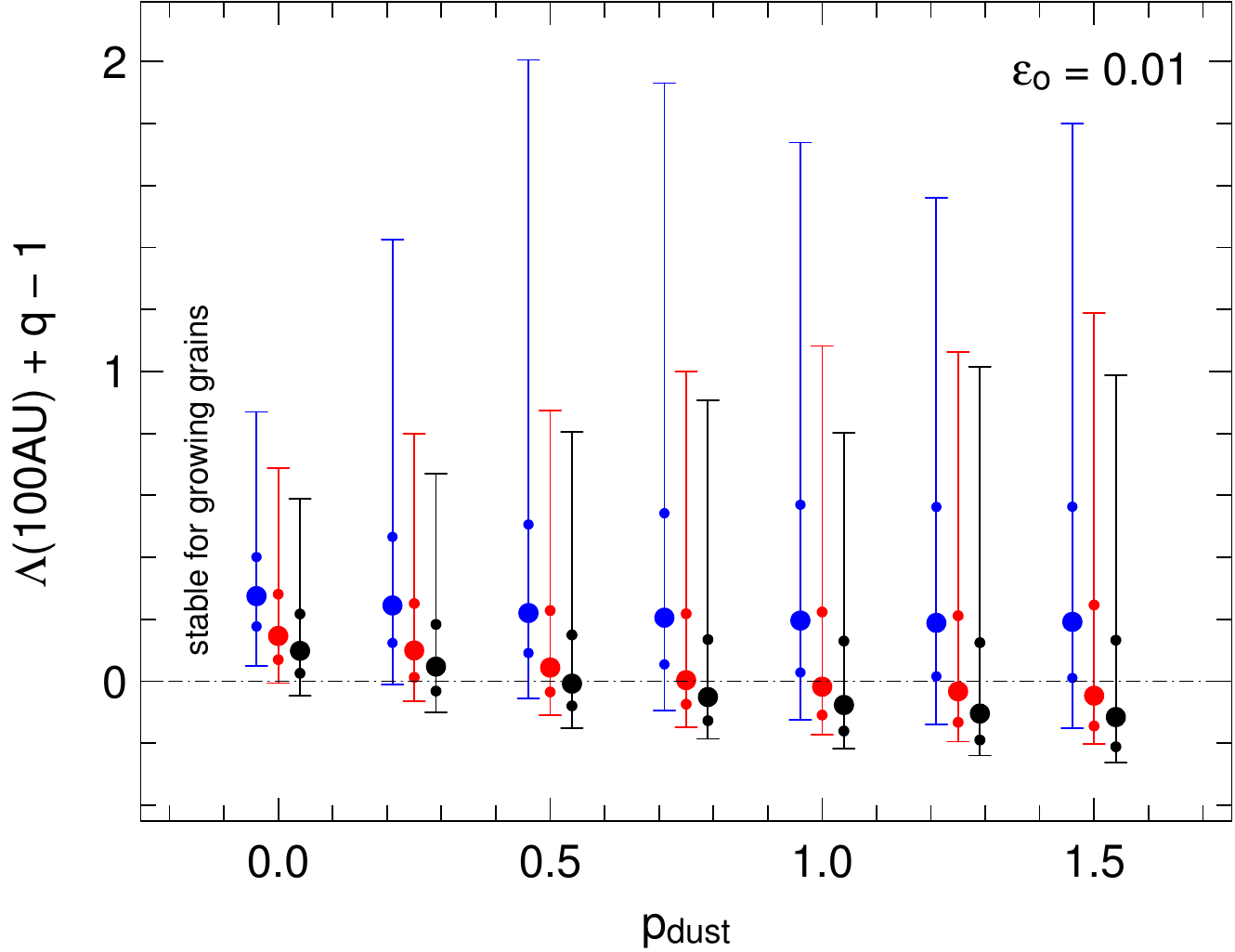}
\caption{
Stability criteria $\Lambda(r) + q - 1 > 0$ for growing dust grains
    when already grown dust grains (left) and vertical settling (right) are present.
    Colour schemes are the same as for Fig.~\ref{fig:grain_size} (left): black:
    all models; red: $a_\mathrm{max} = 1\,\mu$m --$10\,\mu$m, blue; $a_\mathrm{max} = 100\,\mu$m -- $10$\,mm
and for Fig.\ref{fig:settling} (right): no
  settling (black); $\alpha_\mathrm{SS} \geq 10^{-2}$ (low
  settling, red); or
  $\alpha_\mathrm{SS} < 10^{-2}$ (strong settling,
  blue). H(100\,AU) is defined here as $\sqrt{\tfrac{k_B T(r)\, r^3}{G
      M_\mathrm{star}\,
    \mu}}$ where $T(r)$ is the dust midplane temperature, $r = 100$\,AU,
$M_\mathrm{star} = 1\,M_{\sun}$,
and $\mu = 2.3$ is the mean molecular weight.
\label{fig:Lambda_growth_settling}
}
\end{figure*}

The temperature gradient becomes steeper (larger $q$) as the degree
of dust settling increases (smaller $\alpha_\mathrm{SS}$,
Fig.~\ref{fig:settling}).
Dust settling lowers the altitude of the $\tau = 1$
surface, thus reducing the amount of light intercepted at large radii
and steepening the radial temperature profile.
This is critical for discs with a surface density slope close to -1.0,
since dust settling may in that case be enough for $-p+q+1/2$ to become
positive resulting in a strongly increased migration rate for mm/cm
sized grains.

For growing dust grains, two opposite effects have to be considered. Grain
  growth and/or dust settling will steepen the temperature profile, but they will
  also reduce the midplane temperature in the outer disc. As illustrated in
  Fig.~\ref{fig:Lambda_growth_settling} (where results are presented using the local
  hydrostatic scale height), the latter effect dominates: $\Lambda
  + q - 1$ increases in the presence of already grown dust grains and/or
  settling.
Both the presence of grain growth
and vertical settling tend to favour pile-up of the grains.
 In particular, if $a_\mathrm{max} > 100\,\mu$m and
  $\alpha_\mathrm{rm} \leq 10^{-2}$ (blue models in the right panel of  Fig.~\ref{fig:Lambda_growth_settling}), pile-up of dust grains will almost always
  occur, even with $\epsilon_\circ = 0.01$. (87\,\% of this subset of models
  have $\Lambda(100\,\mathrm{AU}) > 1 - q$.)

If the dust opacities are
significantly lowered in the outer disc due to grain growth or dust settling, the outer disc
 structured can be strongly modified, changing from a flared geometry to self-shadowed one
  \citep[e.g.][]{Dullemond04}. Because the outer parts of the disc do not
  intercept much stellar light directly, the temperature drops significantly, resulting
in reduced local sound speed and scale height. Around a T Tauri star, for
instance, the midplane temperature  can drop from 40\,K down to 10\,K at 10\,AU and from 20\,K down to 5\,K at at
100\,AU when
changing the geometry from a flared to a self-shadowed geometry. This translates
into a factor 2 on the gas scale height and the resulting dust-to-gas ratio
required for dust grains to pile up is also reduced by 4. In a self-shadowed
disc with a dust-to-gas ratio of 0.01, growing dust grains will thus stop their
migration at a radius larger than 10\,\% of their initial radius. Such discs
represent excellent sites for forming giant planet cores even in the absence
of dust trapping mechanisms. We
note, however, that discs are immersed in molecular clouds, whose temperatures
range from $\approx$ 7\,K to 15\,K \citep[e.g.][]{Schnee05}. This is likely to
set a lower limit to the temperature reached in the coldest parts of the disc's
midplane and somehow counterbalance
the effects of self-shadowing at very large distances from the star.

\begin{figure*}
\includegraphics[width = 0.31\hsize]{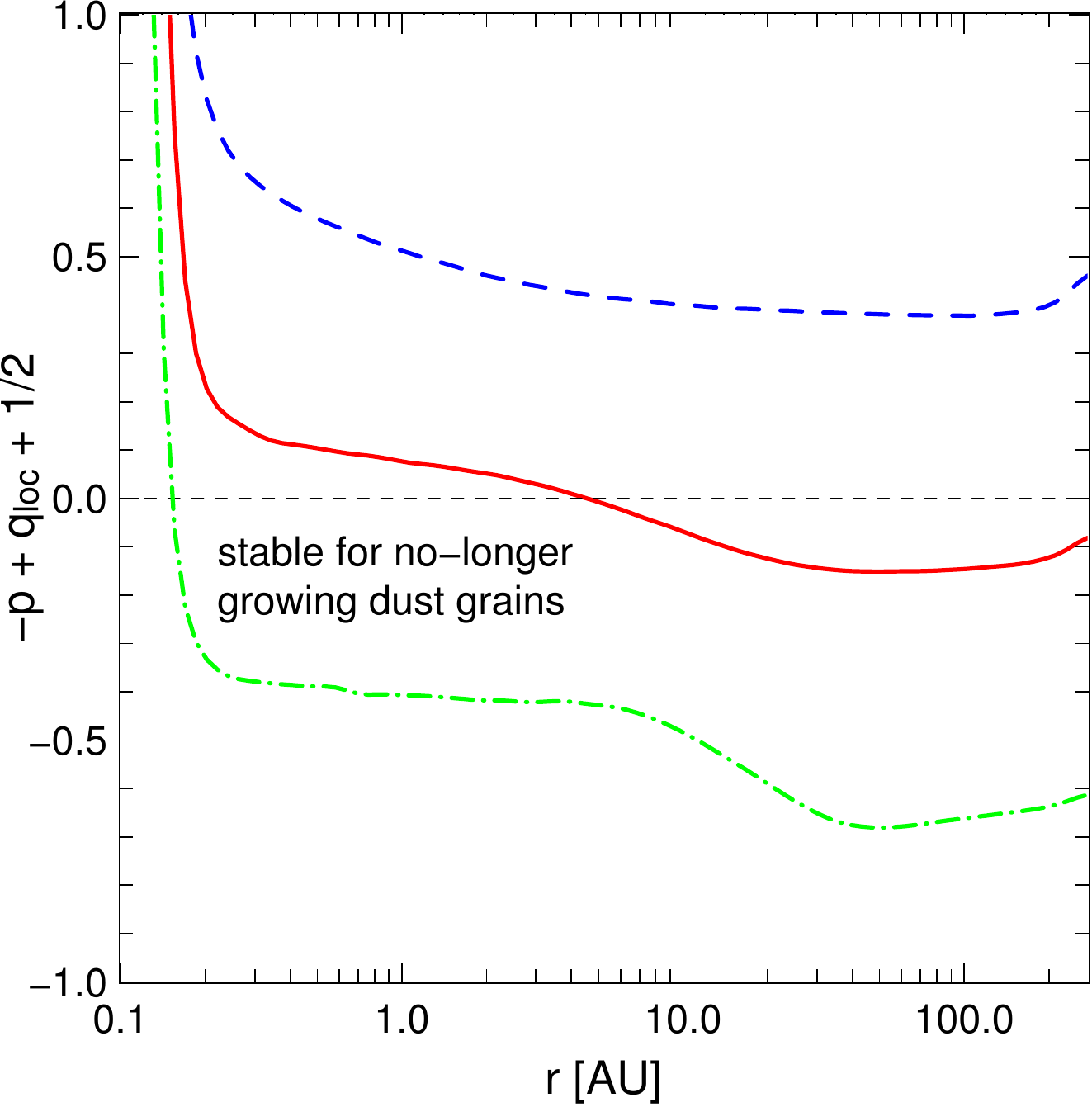}
\hfill
\includegraphics[width = 0.31\hsize]{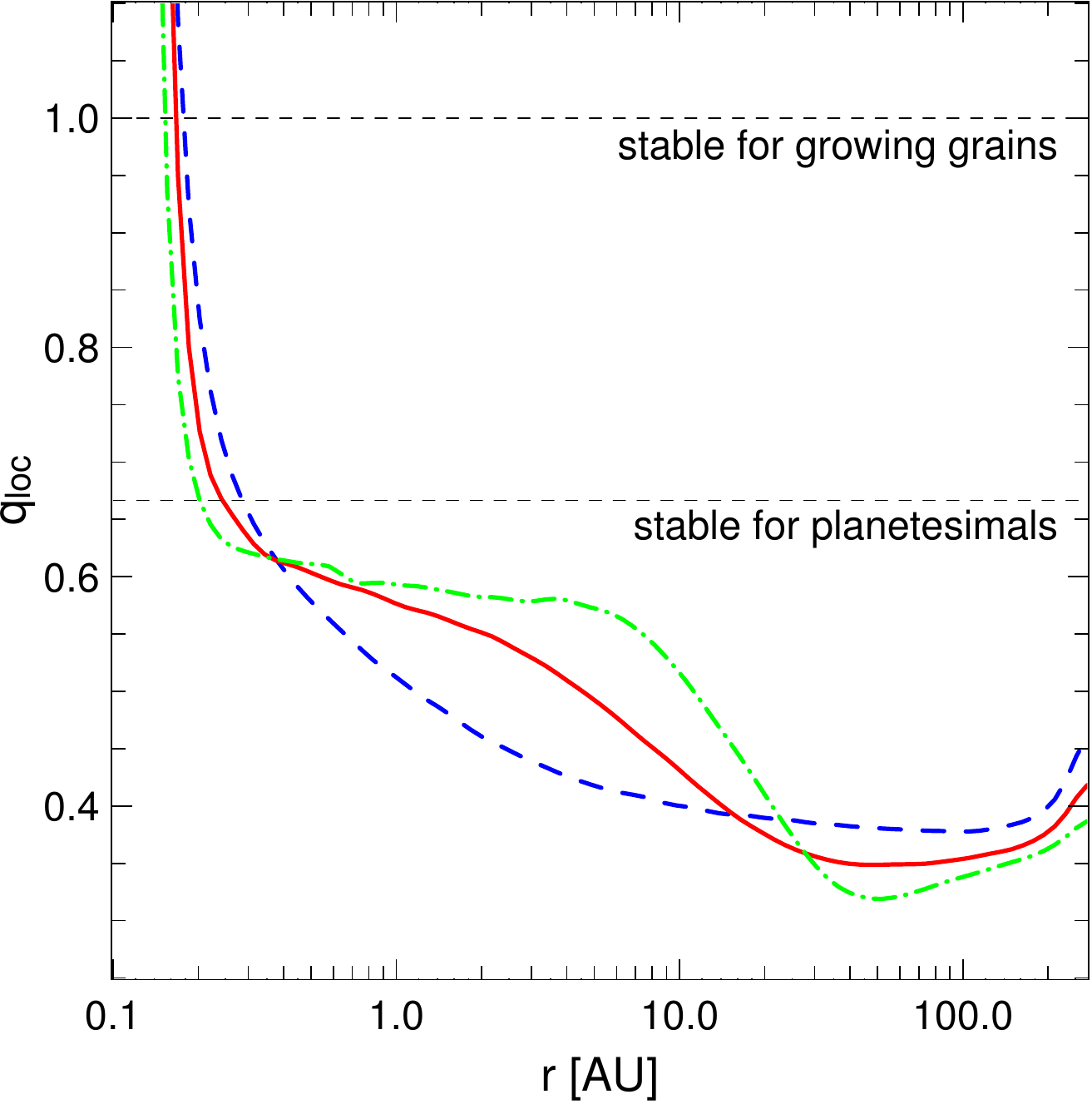}
\hfill
\includegraphics[width = 0.31\hsize]{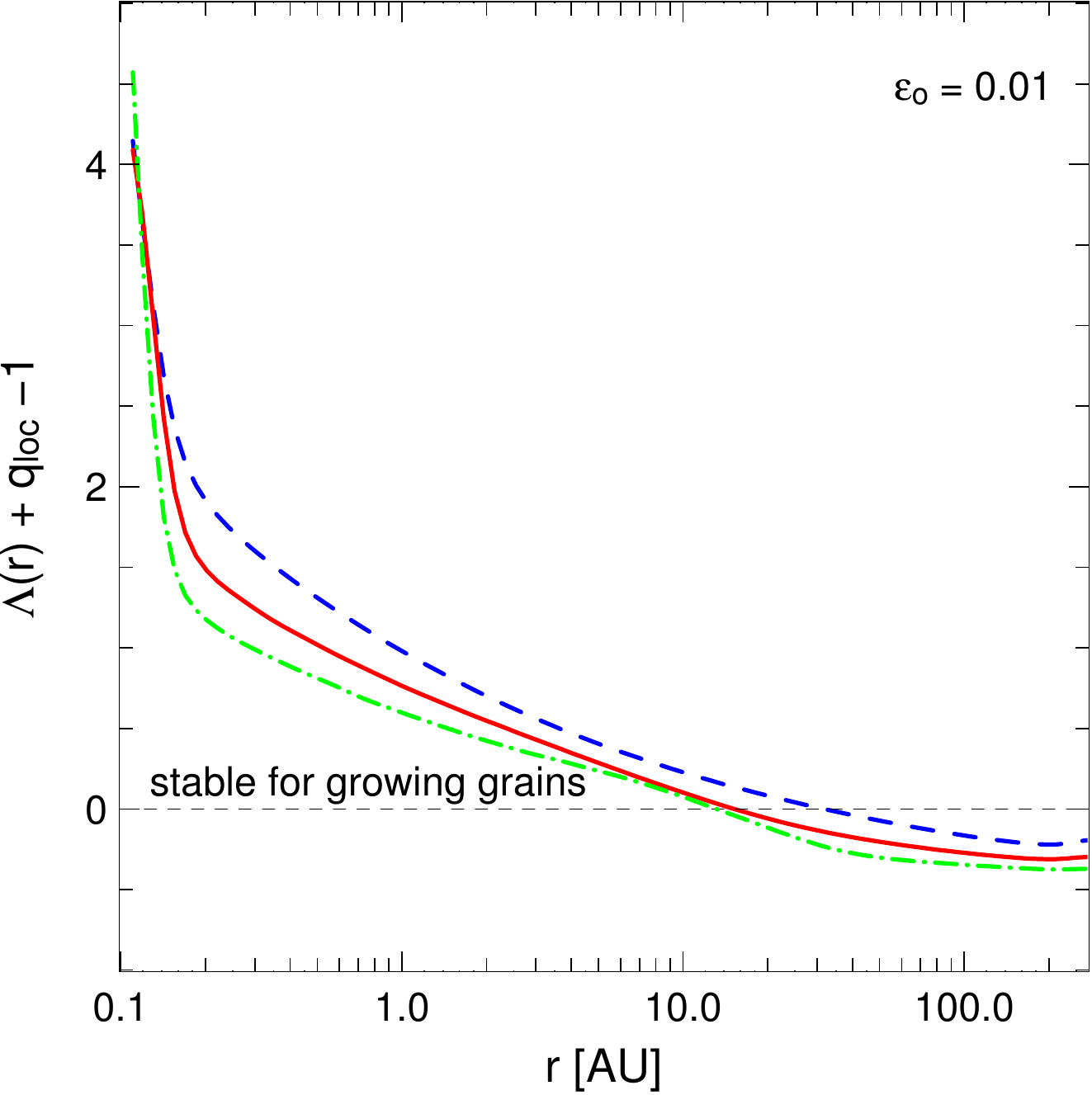}
\caption{Local stability criteria for dust grains (left panel) and
  planetesimals (central panel) and growing dust grains (central and right
  panel) as a function of the radius in the disc, assuming various surface density slopes: $p_{\rm dust}=0.5$ (blue dashed line), $p_{\rm dust}=1$ (red full line), $p_{\rm dust}=1.5$
  (green dash-dotted line). $q_\mathrm{loc}$ is defined as $- \tf{\dd \log
    (T(r))}{\dd \log (r)}$.
 \label{fig:stability_R}}
\end{figure*}

\subsection{The radial-drift barrier in the central few astronomical units}

The midplane temperature can only be approximated by a power law in
the outer regions (where the heating is due to the reprocessing of the
stellar light by the disc's surface).
However, a local temperature slope can be computed at any point to
derive a local stability criteria, describing the proneness of grains
to be accreted across a specific region of the
disc. The stability criteria were only formally established for a power law
temperature distribution, but their evolution as a function of radius, when
the temperature structure differs from a power law, can be used to estimate
if the radial drift will be enhanced or slowed down.

 Figure~\ref{fig:stability_R} shows the stability criteria for
discs of different surface density slopes. In the outer regions, $q_\mathrm{loc}$
remains almost constant (since the temperature structure is
similar to a power law), but with a shallow increase with decreasing
radius.  As a consequence, the conditions for dust stability relative to
    migration remain mostly unchanged over a
    wide range of radii ($> 1 $-- a few 100\,AU).
The $p\approx1$ case is of particular interest since $-p+q+1/2$
can become positive at a relatively large radius ($\approx 10\,$AU).
Inside this radius, migration of mm/cm sized grains becomes
more efficient, and the disc may display a lack of continuum emission at
millimetre wavelengths in the central regions. In contrast, the pile-up of
still growing dust grains becomes more efficient at smaller radius, and the
stability criteria starts to be validated for radii smaller than 15 to 30\,AU
depending on the surface density profile.

The temperature profiles in the very inner disc become extremely steep
($q\gg1$) due to direct heating from the stellar photosphere.
As a
consequence, all dust grains in the Epstein regime, growing or non-growing, as well as
planetesimals in the Stokes regime, will experience very efficient
  migration,
independently of the value of $p$.
Accretion heating (not included in our modelling) can affect the
midplane temperature but only within a few AU of the star
\citep[e.g.][]{DAlessio98} where it can result in an enhanced migration rate for
the dust grains and planetesimals.

\subsection{Long term evolution of the migration process}
\label{sec:discuss_p_dust}

As the dust migrates inward the dust surface density starts to depart
from the initial gas distribution. Neglecting grain growth,
\citet{Youdin2002} show that the drift velocity of grains in the
Epstein regime is given by $v_\mathrm{drift}(r) \propto r^{p_\mathrm{gas} -q
  + 1/2}$. This differential drift velocity results in
dust concentration in the central regions, and steeper dust surface
density profiles if $-p_\mathrm{gas}+q+1/2<0$ (and flatter profiles if
$-p_\mathrm{gas}+q+1/2>0$). More complex profiles are obtained if
grain growth is taken into account (e.g. \citetalias{Laibe08}).

The stability criteria for non-growing grains and planetesimals are only marginally
affected though, because (i) the temperature gradient on large scales
depends moderately on $p_\mathrm{dust}$ (all other parameters being
fixed, see Fig.~\ref{fig:stability_R} right panel for instance) and
(ii) this gradient is
mainly set by the distribution of small grains ($<1\,\mu$m), which
intercept most of the stellar radiation in the disc surface and which
remain coupled to  the gas distribution. We assume here that the small dust
  population is mostly uninfluenced by the growth process.
In the outer disc, where the ``plateau of fast migration'' occurs, the maximum
expected grain size is roughly equal to the optimal size of migration

\begin{equation}
s_{\mathrm{opt}} = \frac{\rho_{\mathrm{g}} c_{\mathrm{s}}}{\rho \Omega_{\mathrm{K}}} ,
\label{eq:sopt}
\end{equation}
where $\rho_{\mathrm{g}}$ and $c_{\mathrm{s}}$ are the gas density and
  sound speed, $\rho$ the intrinsic dust density and $\Omega_{\mathrm{K}}$ the
  Keplerian frequency. The parameter $s_{\mathrm{opt}}$ scales like the disc's
  surface density; \emph{i.e.}, $s_{\mathrm{opt}} \propto
  r^{-p_{\mathrm{gas}}}$. We computed a second version of the grid
  presented in Table~\ref{Tab:param} where the maximum grain size at each radius
  is set $s_{\mathrm{opt}}$, whereas the mass and density slope for each grain
  size bin  is kept unchanged. As expected and as shown in
  Fig.~\ref{fig:pq_migration},the resulting temperature distribution is indeed
  hardly modified.

Assuming that the gas surface density remains more or
less constant (the viscous timescale of the disc being much longer
than the migration timescale), the ability of non-growing dust grains
  and planetesimals to overcome
the radial-drift barrier will thus also only slightly evolve with time.

In the case of growing dust grains, if the condition for pile-up are met,
  the pile up efficiency will increase with time as migrating dust grains will
  enter regions where previous dust grains have already piled-up and where
  the local dust-to-gas ratio is higher (\citetalias{Laibe2014b}).
Interestingly, \citetalias{Laibe08} found that  dust grains break
  through the migration barrier at a few 10\,AU with $p=3/2$, $q=3/4$, and
  $\epsilon_\circ = 0.01$. These simulations were performed with $H/r = 0.05$,
  meaning that dust grains should pile up at large distances for
  $\epsilon_\circ \gtrsim 0.04$. (This number is derived from the right panel of
  Fig.~\ref{fig:dust-to-gas}, scaled by 1/4 to account for the different $H/r$.) A proper
  treatment of hydrodynamics effects results in a pile-up with a smaller
  dust-to-gas ratio illustrating that the criteria we describe are strict
  conditions but that pile-up can occur for less stringent conditions.
 Dust grains that pile up in the inner
    disc regions increase the local dust-to-gas ratio and trigger the pile-up
    of the dust arising from the outer disc (see \citetalias{Laibe2014b}). Moreover, migration
    becomes less efficient as the dust-to-gas ratios reaches values close to
    unity \citep{Nakagawa86}. These two effects favour the pile-up of the
    grains and
    were not accounted for in the \citetalias{Laibe2014b} model. Conversely, pile-up will
  always occur if the  criteria (and the assumptions they are based on)
  are met.

\section{Conclusions}
\label{Sec:conclu}

Based on analytical calculations of the dust evolution in discs (we
  highlight the limitations in section~\ref{sec:limitation}),
we have investigated the effects of a complete treatment of radiative
transfer on the stability of dust grains and planetesimals relative to
radial migration within  protoplanetary discs. We find that for a
significant fraction of the discs we considered, the dust particles
pile up in the disc, potentially providing material for the formation of planetary cores.

Excluding the first few central tenths of AU, all the disc configurations we
explored lead to favourable  temperature profiles ($q<2/3$) for the discs to
retain their planetesimals.

The necessary criterion $q<1$ for discs to retain their
  growing grains is also always fulfilled. In most of the disc configurations,
the conditions are met for dust grains to decouple at a radius larger than
1\,\% if the initial dust-to-gas ratio is higher than 0.03. In the case of a
flat surface density, lower initial dust-to-gas ratios, around 0.01, are
enough. In those cases, dust grains starting at a few hundred AU will
pile up and stop migrating in the regions where telluric planets are formed.
If the initial dust-to-gas ratio is higher, between 0.05 and 0.1, dust grains
pile up at a radius larger than 10\,\% of their initial radius and could
provide the material for forming the cores of massive planets.

Three classes of discs can be distinguished with respect to the outcome of
the radial motion of non-growing grains. Discs with
steep density profiles ($p_{\rm dust}>1.2$) will retain their
grains, whereas dust migration will be very efficient for flat surface
densities ($p_{\rm dust}<0.7$).  For intermediate cases ($0.7<p_{\rm dust}<1.2$),
significant migration of dust grains can occur, but its outcome needs to
be studied via case-by-case detailed simulations coupling
  hydrodynamics and radiative transfer.
Interestingly, this includes the case $p_{\rm dust} \simeq 1$, which corresponds
to the peak of the
distribution of the observed disc profiles.

The presence of large grains and  vertical settling steepens the
temperature profile and tends to
enhance the radial migration of the non-growing grains, favouring accretion
onto the central object. For growing dust grains, however, the reduced
temperature in presence of large grains and/or settling will slow down the
migration and result in pile-up for almost all the disc configurations.

If the settling and/or grain growth increases, the disc
  structure can become self-shadowed, resulting in much lower gas temperature
  and sound speed. In this case, the conditions become much more
  favorable for dust grains piling up and dust-to-gas ratios over around 0.01 are
  enough to stop the migration of the grains coming from 100\,AU at distances over a few AU.

In the central regions of the disc where the stellar radiation can
penetrate directly or accretion heating can contribute, the
temperature profile is so steep ($q\gg1$) that both the dust grains,
whether they are still growing or not,  and
planetesimals are rapidly accreted in all cases. This will also be the
  case if accretion heating becomes  significant and results in
a steep temperature profile in these central regions.

Understanding the detailed outcome of the radial migration in protoplanetary discs requires numerical simulations to catch the complexity of
all the processes at play (\citetalias[e.g.][]{Laibe08}, \citealp{Birnstiel09, Birnstiel10}).

Ideally, these simulations should be coupling radiative transfer and dynamics,
since the conditions for the migration or pile-up of the dust grains are strongly
affected by the thermal structure of the disc.
In this simple study, we performed an exhaustive
  exploration of the parameter space and, based on analytical criteria,
 we give the main results of the
  migration outcome. These results can provide valuable help in tailoring the
  input parameters of the dust evolution codes used to understand the now
  growing evidence of dust radial segregation in discs.

\begin{acknowledgements}
We thank F.~M\'enard, C.~Dougados, J.-C.~Augereau, \mbox{J.-F.~Gonzalez},
S.T.~Maddison, D.J.~Price and J.B.~Kajtar for useful discussions.
C.~Pinte acknowledges funding from the
European Commission's FP7
(contract PERG06-GA-2009-256513) and the
Agence Nationale pour la Recherche of France (contract
ANR-2010-JCJC-0504-01). G.~Laibe is grateful to the Australian
Research Council for funding (contract DP1094585) and acknowledges funding from the European Research Council for the FP7 ERC advanced grant project ECOGAL. Computations
were performed at the Service Commun de Calcul
Intensif de l'Observatoire de Grenoble (SCCI).
\end{acknowledgements}

\bibliographystyle{aa}
\bibliography{biblio}

\begin{appendix}

\section{Validity of the criteria outside the disc midplane}
\label{app:crit3D}

In protoplanetary discs, grains migrate because the local azimuthal velocity of the gas $\vgtheta$ is slightly slower than the Keplerian velocity (\emph{i.e.} sub-Keplerian rotation). Eq.~B.21 of \citetalias{Laibe2012} gives the analytic expression for $\vgtheta$ with respect to the radial and vertical coordinates $(r,z)$:

\begin{equation}
\frac{\vgtheta^{2}}{r} = \frac{\gm}{r^{2}} + \cs^{2} \frac{\mathrm{d}\, \mathrm{ln} \bar{P}}{\mathrm{d}r} + \gm \cs^{2} \int_{0}^{z} \left[\nabla f \times \nabla \cs^{-2}\right].\textbf{e}_{\theta} \mathrm{d}z' .
\label{eq:vthetaelegant}
\end{equation}
where $\bar{P}$ is the gas pressure in the disc midplane, $\cs (r,z)$ is the sound speed and $f$ is a function defined by
\begin{equation}
f\left(r,z\right) = \frac{1}{r} - \frac{1}{\sqrt{r^{2}+z^{2}}} .
\label{eq:deff}
\end{equation}

The first term of the right-hand side of Eq.~\ref{eq:vthetaelegant} is the
Keplerian contribution. The second term is the pressure gradient term, which
depends only on $r$. \citetalias{Laibe2012}'s criteria originate from this contribution. The third term is the baroclinic term. It contributes to the deviation to the Keplerian velocity only if $z \ne 0$. However, the ratio between the baroclinic term and the pressure gradient term is of order $\mathcal{O}\left(\left(z/H \right)^{2} \right )$. This implies that for particles slightly outside the disc midplane, the baroclinic term has a negligible contribution and all the criteria used above apply.

\section{Models including simple dust migration}

\begin{figure}
\includegraphics[width = \hsize]{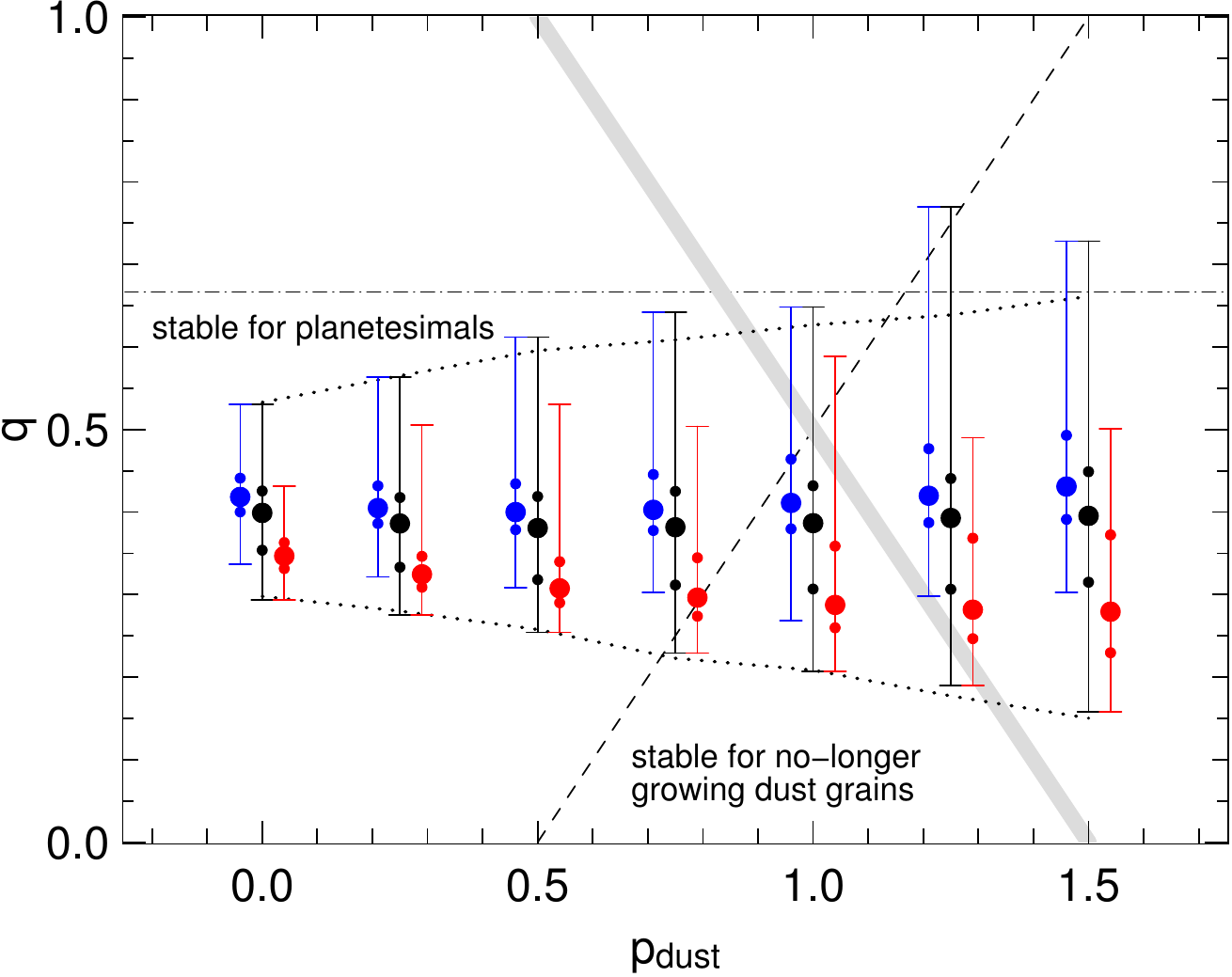}
\caption{Same as Fig.~\ref{fig:grain_size} but with migration for $s >
  s_\mathrm{opt}$. Taking into account the migration only marginally affects
  the temperature profile and subsequent migration of the remaining
  grains. The 2 dotted lines show the envelope of the models without
    migration (displayed in Fig.~\ref{fig:grain_size}) for comparison.
\label{fig:pq_migration}}
\end{figure}

The results of the grid of models where we assume that $a_\mathrm{max}$ is a
function of the radius and is set to be equal to the optimum size of migration~:
\begin{equation}
a_\mathrm{max} = s_{\mathrm{opt}} = \frac{\rho_{\mathrm{g}} c_{\mathrm{s}}}{\rho \Omega_{\mathrm{K}}} ,
\end{equation}
are presented in Fig.~\ref{fig:pq_migration}.

\end{appendix}

\end{document}